\documentstyle[preprint,pre,aps]{revtex}
\newcounter{saveeqn}

\title{Multi-State Image Restoration by Transmission of
Bit-Decomposed Data} 
\author{Takashi Tadaki} 
\address{Department of Physics, Tokyo Institute of Technology,
Oh-okayama, Meguro-ku, Tokyo 152-8551, Japan}
\author{Jun-ichi Inoue}
\address{Complex Systems Engineering, Graduate School of
Engineering, Hokkaido University, 
N13-W8, Kita-ku, Sapporo 8628, Japan}
\date{\today}
\begin{document}
\maketitle
\thispagestyle{empty}
\begin{abstract}
We report on the restoration of gray-scale image 
when it is decomposed into a binary form before transmission. 
We assume that a gray-scale image expressed by a set of $Q$-Ising spins is
first decomposed into an expression using Ising (binary) spins by means
of the threshold division, namely, we produce ($Q-1$) binary Ising spins from a
$Q$-Ising spin by the function $F(\sigma_{i}-m) = 1$ if the input data
$\sigma_{i} \in \{0,\cdots,Q-1\}$ is $\sigma_{i} \geq m$ and 0
otherwise, where $m$ $\in\{1,\cdots,Q-1\}$ is the threshold value.
The effects of noise are different from the case where the raw $Q$-Ising
 values are sent. We investigate which is more effective to use the binary
 data for transmission or to send the raw $Q$-Ising values. 
By using the mean-filed model, 
we analyze the performance of our method quantitatively. 
In order to investigate what kind of original picture is efficiently
 restored by our method, the standard image in two dimensions is 
 simulated by the mean-field annealing, and we compare the performance
 of our method with that using the $Q$-Ising form. 
We show that our method is more efficient than the one using the
 $Q$-Ising form when the original picture has large parts in which the
 nearest neighboring pixels take close values. 
\end{abstract}
\mbox{}
\clearpage
\setcounter{page}{1} 
\section{Introduction}
Statistical mechanical approaches to the problems in information science
are employed frequently, and it is now known that such methods are very
useful in various problems \cite{bib0}. 
Among such problems, 
the image restoration problem has been investigated both theoretically 
and practically \cite{bib4,bib5,bib6}. 
We usually send and receive the information
 by means of various networks. The information is the data of document,
 sound, picture, and so on. However, it is generally impossible for a
 receiver to receive complete transmitted data when a sender transmits
 something, because the data are transmitted through a noisy channel. If
 we restrict the type of data to that of the picture, the original image is
 affected by some kind of noise when it is sent by a defective fax, a fickle
 e-mail, etc. When we receive such a corrupted image, we have to
 convert it using some kind of filter to obtain the original image.
Basically it is the image restoration problem to estimate the original
 data (the original image) from the received, corrupted data (the
 degraded image).
We may regard the digital picture as a discrete spin system.
For example, a black and white image (WBI) corresponds to an
Ising spin system by identifying the white color with $+1$ and the
black with $-1$. Furthermore, the theory of image restoration
is constructed by considering that two axes of the plane, $x$-axis and
$y$-axis, correspond to two axes of time in the stochastic process. 
That is, the Markov process that the event occurs at a specific time is
affected by what happened at neighboring pixels at a one-time step before.

There are mainly two standard approaches to the image restoration by means
of the method of statistical mechanics. One is called the maximum a
posteriori (MAP) estimation in which the estimation of the original
image is given by maximizing a posterior probability distribution.
This estimation will be seen to correspond to a search of the ground 
state of spin system described by the effective Hamiltonian in the 
context of statistical mechanics.
Another is the estimation in which we regard the expectation value with
respect to the maximized marginal posterior probability at each site in
 thermal equilibrium as the original image and is called the maximum
 posteriori marginal (MPM) estimation. This estimation is also called
 the finite temperature restoration. Therefore, the MPM estimation includes 
the MAP estimation. Among the two estimations, it was proposed by Marroquin
 et al \cite{bib14} that the MPM estimation gives better performance than the
MAP estimation. Nishimori and Wong \cite{bib15} proved this fact
 for black and white images using a rigorous inequality. With respect
 to the gray-scale image, the same has been shown by
 Tanaka \cite{bib16} from a different viewpoint.

Not only black and white images but also gray scale images are actively
investigated by many people in the field of statistical mechanics. 
Restoration of gray scale image using chiral Potts spin 
\cite{bib20} may not be appropriate since chiral Potts spin can not 
express the distance among different states.
However $Q$-Ising spin \cite{bib19} can express the gray scale level at least.
Restoration of gray scale image using $Q$-Ising spin was
first investigated by Inoue and Carlucci \cite{bib21}.

Inspired by their studies, 
we investigate the restoration of the gray scale image expressed
by the $Q$-Ising spin when it is decomposed into the Ising spin before
transmission. In this method, the gray scale image is decomposed into
binary data by means of the method of threshold division before it is
transmitted. Then the bit-decomposed data is transmitted through a noisy
channel. Therefore the effects of noise is different from those in the 
$Q$-Ising form. We show how well the original image is restored in our
method in comparison to the restoration using the $Q$-Ising form.

This paper is organized as follows. In the next section, we explain the
general formulation of image restoration and the method of threshold
division. In Sec. III, we analyze the static 
properties of image restoration in the infinite-range model.
In Sec. IV, we verify the result of the infinite-range model in
realistic pictures by means of Monte Carlo simulation.
In Sec. V, standard image is restored by the method of mean-field
annealing. The performance of restoration in our method is 
compared with that using the $Q$-Ising form.
The final section is denoted to summary and discussions.

\section{General formulation}
\subsection{Image restoration}
A gray scale image is represented by a set $\{\xi\}_{i=1,\cdots,N}$ of
fixed values. The variable $\xi_{i}$ takes an integer value between $0$ and
$Q-1$. Even in black and white image in addition to gray scale
image, most of natural images have nontrivial structures generally.
Therefore, it is impossible to discuss the property
of a given specific natural image exactly. Nevertheless we may notice that an
eminent property of natural images is local smoothness.
Therefore we assume that the original image is generated by the
Boltzmann probability represented by the following: 
\begin{equation}
P_{s}(\{\xi\})=
\frac{1}{{\cal Z}(\beta_{s})}
{\exp}\left[
-\frac{\beta_{s}}{2z}\sum_{(ij)}(\xi_{i}-\xi_{j})^{2}
\right],
\label{ir1}
\end{equation}
where $(ij)$ represents interacting sites and $z$ is the coordination
number. $Z(\beta_{s})$ is the normalization constant, and
$\beta_{s}(=T_{s}^{-1})$ is the inverse temperature to generate the
original image. If $\beta_{s}$ is large, an original image with many
clusters consisting of the same value is generated.

A degraded image is generated by sending data of the original image
through a noisy channel. We consider two kinds of noise.  
One is the binary noise caused by a binary symmetric channel
(BSC) and another is the Gaussian noise caused by the Gaussian channel
(GC).

\subsubsection{Gaussian channel}
In the Gaussian channel, the output $\tau_{i}$ for an input $\xi_{i}$ is
a Gaussian random variable with mean $\tau_{0}\xi_{i}$ and variance
$\tau^{2}$. The probability distribution of output given the input
$\{\xi\}$ is written as 
\begin{equation}
P(\{\tau\}|\{\xi\})=\frac{1}{\sqrt{2\pi}\tau}{\exp}
\left[-\frac{1}{2\tau^{2}}
\sum_{i}(\tau_{i}-\tau_{0}\xi_{i})^{2}
\right].
\label{ir2}
\end{equation}
The degraded image $\{\tau\}_{i=1,\cdots,N}$ is generated by this
probability distribution.

According to the Bayes formula, the posterior probability
$P(\{\sigma\}|\{\tau\})$ that the estimate of source sequence,
namely the restored image, is $\{\sigma\}_{i=1,\cdots,N}$, provided that
the output is $\{\tau\}$, is given as 
\begin{eqnarray}
P(\{\sigma\}|\{\tau\}) & = & \frac{P(\{\tau\}|\{\sigma\})P_{m}(\{\sigma\})}
{{\rm tr}_{\{\sigma\}}P(\{\tau\}|\{\sigma\})P_{m}(\{\sigma\})} \nonumber \\
\mbox{} & \sim &
\exp {\Bigg (} -h\sum_{i}(\sigma_{i}-\tau_{i})^{2}-
\frac{\beta_{m}}{2z}\sum_{ij}(\sigma_{i}-\sigma_{j})^{2}
{\Bigg )}\nonumber \\
& \equiv &
\exp(-{\cal H}_{{\rm G}}).
\label{ir3}
\end{eqnarray}

Since we can use the degraded image only and do not know the other
information of original image, we introduced the model prior 
$P_{m}(\{\sigma\})$ to represent a priori knowledge on natural image: 
$P_{s}(\{\xi\})$ represented by equation (\ref{ir1}).
Furthermore prior parameters $\beta_{s}$ and ${\tau_{0}}/{\tau^{2}}$
to control the generation of original image and degraded image,
respectively, are unknown quantities.
Accordingly, we have to use the so-called hyper-parameters $\beta_{m}$ and
$h$ instead of prior parameters $\beta_{s}$ and
${\tau_{0}}/{\tau^{2}}$.
Then, controlling these hyper-parameters, we can obtain the optimal
restored image.

\subsubsection{Binary symmetric channel}
When binary data $\{\xi_{b}\}\,(Q=2)$ takes $0$ or $1$, the type of
noise can also be binary. In such a case, a pixel of the
original image is flipped with the probability $p$, the error rate.
The error probabilities of flipping the signal +1 to 0
and 0 to +1 are the same. The probability distribution of this output
$\{\tau_{b}\} \in \{0,1\}$ is expressed as follows.
\begin{equation}
P(\{\tau_{b}\}|\{\xi_{b}\})=
\frac{1}{Z_{\tau}}{\exp}\left[
-\beta_{\tau}\sum_{i}(\tau_{i,b}-\xi_{i,b})^{2}
\right],
\label{ir4}
\end{equation}
where
\begin{eqnarray}
Z_{\tau} &=& {\rm Tr}_{\tau_{b}}{\exp}\left[
-\beta_{\tau}\sum_{i}(\tau_{i,b}-\xi_{i,b})^{2}\right] \nonumber \\
& = & \left[ 1+{\rm e}^{\beta_{\tau}}\right]^{N}. \nonumber
\label{ir5}
\end{eqnarray}
The parameter $\beta_{\tau}$ is defined by
$p$ becomes
\begin{eqnarray}
\beta_{\tau}=\ln \frac{1-p}{p}   .
\label{ir6}
\end{eqnarray}
The posterior probability in this case is 
\begin{eqnarray}
P(\{\sigma_{b}\}|\{\tau_{b}\}) & = &
\frac{P(\{\tau_{b}\}|
\{\sigma_{b}\})
P_{m}(\{\sigma_{b}\})}
{{\rm tr}_{\sigma_{b}}P(\{\tau_{b}\}|
\{\sigma_{b}\})
P_{m}(\{\sigma_{b}\})} \nonumber \\
\mbox{}& \sim & 
\exp{\Bigg (}-h\sum_{i}(\sigma_{i,b}-\tau_{i,b})^{2}-
\frac{\beta_{m}}{2z}\sum_{ij}(\sigma_{i,b}-\sigma_{j,b})^{2}{\Bigg )}\nonumber \\
&\equiv&
\exp(-{\cal H}_{{\rm B}}).
\label{ir7}
\end{eqnarray}
\subsection{MPM estimation and mean square error}
Next, we discuss the method to estimate the original image from the preceding
posterior probability distribution and to evaluate the restored image
which is obtained by such estimations.

We consider the marginal distribution obtained from the posterior
probability
distribution equation (\ref{ir3}) (or (\ref{ir7})) to estimate the original
image:
\begin{eqnarray}
\bar{P}(\sigma_{i}|\{\tau\}) & \equiv &
\sum_{\sigma\neq \sigma_{i}}
P(\{\mbox{\boldmath $\sigma$}\}|\{\mbox{\boldmath $\tau$}\})
\label{ir8}
\end{eqnarray}
Using this marginal probability distribution, we calculate the local
magnetization at a site $i$ as given by
\begin{eqnarray}
\langle \sigma_{i} \rangle_{\beta_{m},h} & \equiv  &
\sum_{\sigma_{i}=0}^{Q-1}\sigma_{i}\bar{P}
(\sigma_{i}|\{\tau\}) .
\label{ir9}
\end{eqnarray}
Since the above expectation value is a continuous real number, we need
to change it into an integer by means of the following function
$\Omega$: because the original image consists of the discrete value in
digital images, the restored image is the estimation of original image
should, properly, take the discrete value. 
\begin{eqnarray}
\Omega(\langle \sigma_{i}\rangle_{\beta_{m},h})&\equiv&
\sum_{k=1}^{Q}k{\Bigg[}\Theta{\bigg(}\langle
\sigma_{i}\rangle_{\beta_{m},h} - \frac{2k-1}{2}{\bigg)}-
\Theta{\bigg(}\langle \sigma_{i}\rangle_{\beta_{m},h} -
\frac{2k+1}{2}{\bigg)}{\Bigg]}
\label{ir10}
\end{eqnarray}
A real number $\langle \sigma_{i}
\rangle_{\beta_{m},h}$ is translated into the closest integer.
Then we regard the above discrete value $\Omega(\langle \sigma_{i}
\rangle)$ as the value of $i$-th pixel for restored image at finite
$\beta_{m}$ and $h$ in the finite temperature process.\\
In the MAP estimation, the estimation of the original image is regarded
as the set $\{\sigma_{i}\}$
which maximizes the posterior probability distribution (Eq.s 
(\ref{ir3}),(\ref{ir7})). In other words, it is the set that minimizes the
energy of the system described by the Hamiltonian ${\cal H}_{{\rm G}}$ 
or ${\cal H}_{{\rm B}}$ and is actually the ground state. 
Consequently, the above
estimation corresponds to the MAP estimation when $T_{m} \rightarrow 0$
$(\beta_{m}\rightarrow \infty)$ keeping $H=h/\beta_{m}$ constant.

It is very important to evaluate the performance of restoration by means
of these estimations. For this purpose, we assume that we know the
original image and evaluate the performance of restoration by measuring
the distance between the original and the restored images
given by the above estimation. In order to measure the distance between
the original and the restored images, we use the following mean
square error as the distance, 
\begin{eqnarray}
H_{\rm D}
 =  \frac{1}{N} \sum_{i} [\xi_{i} - {\Omega}(\langle \sigma_{i}
\rangle_{\beta_{m},h})]^{2},
\label{ir11}
\end{eqnarray}
whose value depends on the hyper-parameters $h$ and $\beta_{m}$
appearing in the thermal average.

\subsection{Method of threshold division}
We next discuss the restoration of gray-scale image using
bit-decomposed data. It is expected that the effects of noise on the
binary data is lower than that on the $Q$-value data for image restoration:
the binary data is estimated easily compared with the $Q$-value data
because of the explicit representation.
Since the original image is estimated by using the information of
degraded image, the performance of restoration
ought to be deeply affected by the effects of noise. 
Therefore we expect that the performance of restoration is improved by
using the binary data instead of the $Q$-value data.

We generate binary data from $Q$-value data using the $Q$-Ising form
by means of the following function 
\begin{equation}
\Theta (\xi_{i} - k) = \xi_{i,k} = 
1 \,\, \mbox{$(\xi_{i}\geq k)$}, \,\, 
0 \,\, \mbox{$(\xi_{i}< k)$} 
\end{equation}
where $k\,(\in \{1,2,\cdots,Q-1\})$ is the threshold value and
$\xi_{i}\,(\in \{0,1,\cdots,Q-1\})$ is the input data. We obtain
$(Q-1)$ sets of binary data from a $Q$-value data by using this function
with threshold value changed from 1 to $Q-1$. This method is called the
{\it threshold division}. For example, we consider a set with $Q=3$ ($0$, or
$1$, or $2$) and six pixels: $\{1,2,1,1,0,2\}$. If we insert this set into
the above function with the threshold value $k=2$, the binary set
$\{0,1,0,0,0,1\}$ is generated.
The set $\{1,1,1,1,0,1\}$ is also generated when $k=1$.
Thus two binary sets $\{0,1,0,0,0,1\}$ and
$\{1,1,1,1,0,1\}$ are generated from the $3$-value set
$\{1,2,1,1,0,2\}$. These operations are depicted in FIG. \ref{fig2}.
The expression of binary data $\xi_{i,k}(\in \{0,1\})$ produced by this
method is different from the usual binary notation because the relation between
a $Q$-value data and ($Q-1$) sets of binary data becomes 
\begin{equation}
 \xi_{i}=\xi_{i,1}+\xi_{i,2}+\cdots+\xi_{i,Q-1}=\sum_{k=1}^{Q-1}\xi_{i,k} .
\label{thresh1}
\end{equation}
We call binary data generated by this 
threshold division the {\it bit-decomposed data} (BDD).

Next, the process in the restoration of gray-scale image using the
bit-decomposed data becomes the following.
\begin{enumerate}
\item
Decompose a $Q$-value data (original image) into ($Q-1$) sets of binary
     data before transmission.
\item
Send ($Q-1$) sets of binary data through a noisy channel.
\item
Receive ($Q-1$) sets of corrupted binary data (degraded image).
\item
Restore the original image from the degraded image.
\end{enumerate}
Obviously, this procedure is different from 
the method in which the raw $Q$-value data
are
sent.

Following the general formulation mentioned in the previous section, we
formulate the above process. The posterior probability necessary for
restoration is the following
\begin{eqnarray}
P(\{\sigma\}|\{\tau_{1}\},\{\tau_{2}\},
\cdots \{\tau_{Q-1}\}) & = &
\frac{P(\{\tau_{1}\},\{\tau_{2}\},\cdots \{\tau_{Q-1}\}|\{\sigma\})
P_{m}(\{\sigma\})}{{\rm tr}_{\{\sigma\}}
P(\{\tau_{1}\},\{\tau_{2}\},\cdots \{\tau_{Q-1}\}|\{\sigma\})
P_{m}(\{\sigma\})} \nonumber \\
& \sim &
{\exp}\left(
-h\sum_{i}\sum_{k}(\sigma_{i,k}-\tau_{i,k})^{2}-
\frac{\beta_{m}}{2z}\sum_{i,j}(\sigma_{i}-\sigma_{j})^{2}
\right),
\label{thresh2}
\end{eqnarray}
where we used $\{\sigma\}$ to denote the dynamical variables used to
 estimate $\{\xi\}$. 
The notation $\sigma_{i,k}\in \{0,1\}$ is the estimation of
 decomposed data $\xi_{i,k}$. The notation $\tau_{i,k} \in \{0,1\}$ is
the degraded binary data. 
The relation between $\sigma_{i,k},\tau_{i,k}$ and
$\sigma_{i},\tau_{i}$ is the same as that between $\xi_{i}$ and
$\xi_{i,k}$ (Eq. (\ref{thresh1})).\\
In the restoration of gray-scale image using the $Q$-Ising form, the
posterior probability distribution is given by
\begin{eqnarray}
P(\{\sigma\}|\{\tau\}) &\sim&
{\exp}\left(
-h\sum_{i}(\sigma_{i}-\tau_{i})^{2}-
\frac{\beta_{m}}{2z}\sum_{ij}(\sigma_{i}-\sigma_{j})^{2}
\right).
\label{thresh3}
\end{eqnarray}
The difference of the posterior probability between our method
(\ref{thresh2}) and the $Q$-Ising form (\ref{thresh3}) is only in the
random field term. That is to say, the
effects of noise in our method are different from the case where the raw
$Q$-Ising data are sent.
\section{Analysis of the infinite-range model}
In this section, we discuss the restoration of the gray-scale 
image in the infinite-range model (mean-field model).
Since the infinite-range model 
is the model in which all pixels
interact mutually, it is not useful for restoration of real
images directly. However, the analysis of image restoration in the
infinite-range model is very useful in understanding
qualitatively the property of macroscopic quantities. 
For this reason, 
we investigate the averaged performance of our method 
by using the infinite range model. 

Suppose that the  prior probability generates the original 
image in the infinite-range model like Eq. (\ref{ir1}).
Then the $i$-th pixel interacts  all other pixels
and the prior probability distribution is represented as
\begin{equation}
P_{s}(\{\xi\})=
\frac{1}{Z(\beta_{s})}
{\exp}\left[
-\frac{\beta_{s}}{2N}\sum_{ij}(\xi_{i}-\xi_{j})^{2}
\right],
\label{stat1}
\end{equation}
where $N$ is the system size and $Z(\beta_{s})$ is the partition function
of the ferromagnetic $Q$-Ising model.
We assume that the original image is generated by this probability both
in the bit-decomposed data case and the $Q$-Ising case. In this section
we adopt as the original image a snap shot of the system produced by the
above probability.

We treat the Gaussian channel for simplicity. In our method, the
Gaussian channel in which input data are expressed 
in a binary form is given by
\begin{eqnarray}
P_{k}(\{\tau_{i,k}\}|\{\xi_{i,k}\}) & = &
\frac{1}{\sqrt{2\pi}\tau}{\exp}\left[-\frac{1}{2{\tau}^{2}}\sum_{i}
(\tau_{i,k}-\tau_{0}\xi_{i,k})^{2}\right],\nonumber \\
& &
\label{stat2}
\end{eqnarray}
where $\tau_{i,k}$ is a Gaussian random variable with mean
$\tau_{0}\xi_{i,k}$ and variance $\tau^{2}$. 
The input $\xi_{i,k}$ takes $0$ or $1$.
For comparison, the $Q$-Ising case is denoted
\begin{eqnarray}
P(\{\tau_{i}\}|\{\xi_{i}\}) =
\frac{1}{\sqrt{2\pi}\tau'}{\exp}\left[
-\frac{1}{2{\tau'}^{2}}\sum_{i}(\tau_{i}-\tau'_{0}\xi_{i})^{2}
\right],
\label{stat3}
\end{eqnarray}
where $\xi_{i}$ takes a value between $0$ and $Q-1$, and $\tau_{i}$ is a
Gaussian random variable with mean $\tau'_{0}\xi_{i}$ and variance
${\tau'}^{2}$. 
We calculate the posterior probability using Eq. (\ref{stat2}) and
Eq.  (\ref{stat3}). \\
In the bit-decomposed data case;
\begin{eqnarray}
P(\{\sigma\}|\{\tau_{1}\},
\{\tau_{2}\},\cdots \{\tau_{Q-1}\})
& \sim &
{\prod_{k}^{Q-1}P_{k}(\{\tau_{k}\}|\{\sigma_{k}\})
P_{m}(\{\sigma\})} \nonumber \\
& \sim &
{\exp}\left(
-h\sum_{i}\sum_{k}(\sigma_{i,k}-\tau_{i,k})^{2}-
\frac{\beta_{m}}{2N}\sum_{i,j}(\sigma_{i}-\sigma_{j})^{2}
\right) \nonumber \\
&\equiv &
{\exp}\left( - H_{{\rm eff}} \right).
\label{stat4}
\end{eqnarray}
In the $Q$-Ising case;
\begin{equation}
P(\{\sigma\}|\{\tau\})
 \sim
{\exp}\left(
-h\sum_{i}(\sigma_{i}-\tau_{i})^{2}-
\frac{\beta_{m}}{2N}\sum_{i,j}(\sigma_{i}-\sigma_{j})^{2}
\right).
\label{stat5}
\end{equation}
The difference of these expressions and equations (\ref{thresh2}),
(\ref{thresh3}) in previous section is only that the coordination number
$z$ became $N$.

We calculate the free energy from these probability
distributions to clarify the behavior of macroscopic quantities.
Since there is randomness in the field for the spin system
described by the effective Hamiltonian $H_{{\rm eff}}$ in equations
(\ref{stat4}) and (\ref{stat5}), the free energy must be averaged over
the probability distribution of the random field in addition to the
thermal average. 
Accordingly, we calculate the free energy per site by
\begin{equation}
f = F/N = -\frac{1}{N\beta_{m}}[\ln Z].
\label{stat6}
\end{equation}
By using the replica method, 
we obtain the following 
replicated partition function,
\begin{eqnarray}
[Z^{n}] &=& \sum_{\xi}\sum_{\tau}P_{s}(\{\xi\})P(\{\tau\}|\{\xi\})
{\rm Tr}_{\sigma}{\rm e}^{-H_{{\rm eff}}^{{\rm rep}}} \nonumber \\
& = &
{\rm Tr}_{\xi}\int \prod_{i,k}d\tau_{i}^{(k)}
\frac{1}{(\sqrt{2\pi}\tau)^{(Q-1)N}}\exp[-\frac{1}{2\tau^{2}}
\sum_{i}(\tau_{i}-\tau_{0}\xi_{i})^{2}]
\times \exp[-\frac{\beta_{s}}{2N}\sum_{ij}(\xi_{i}-\xi_{j})^{2}] \nonumber\\
& &
\times {\rm Tr}_{\sigma}
\exp[-h\sum_{i}\sum_{k}\sum_{\alpha}(\tau_{i,k}-\sigma_{i,k}^{\alpha})^{2}-
\frac{\beta_{m}}{2N}\sum_{ij}\sum_{\alpha}
(\sigma_{i}^{\alpha}-\sigma_{j}^{\alpha})^{2}],
\label{stat8}
\end{eqnarray}
where $\alpha$ is the label of $n$ dummy replicas.
Assuming replica symmetric ansatz $m_{\alpha}=m\quad (\forall \alpha)$,
the following expressions of the order parameters are derived. 
\begin{equation}
[\xi_{i}]=m_{0}=\frac{1}{Z(\beta_{s})}{\rm tr}_{\xi}
\xi {\rm e}^{2m_{0}\beta_{s}\xi- \beta_{s}\xi^{2}},
\label{stat9}
\end{equation}
\begin{equation}
[\langle \sigma_{i}^{\alpha}\rangle ]=m=\frac{1}{Z(\beta_{s})}{\rm
tr}_{\xi}{\rm e}^{2m_{0}\beta_{s}\xi - \beta_{s}\xi^{2}}\int Du_{1} \cdots \int
Du_{Q-1}
\frac{{\rm tr}_{\sigma} \sigma\exp(\tilde{H}_{{\rm eff}})}
{{\rm tr}_{\sigma}\exp(\tilde{H}_{{\rm eff}})},
\label{stat10}
\end{equation}
where we defined 
$Du \equiv ({\rm d}u/\sqrt{2\pi}){\rm e}^{-u^{2}/2}$  and  
\begin{equation}
\tilde{H}_{{\rm eff}} \equiv 2\beta_{m}m\sigma - \beta_{m}\sigma^{2} -
 h\sum_{k}\sigma_{k}^{2}+
2h\tau\sum_{k}u_{k}\sigma_{k}+2h\tau_{0}\sum_{k}\xi_{k}\sigma_{k}.
\label{stat13}
\end{equation}
Equation (\ref{stat9}) represents the magnetization of the original image in
the infinite range model. The behavior of the source magnetization $m_{0}$
as a function of the temperature $\beta_{s}$ is shown in FIG. \ref{fig4}.
Equation (\ref{stat10}) is the equation of state for the magnetization
of the restored image. In the limit $N \rightarrow \infty$, the mean
square error (Eq. (\ref{ir11})) is  rewritten as
\begin{equation}
H_{{\rm D}} = \frac{1}{Z(\beta_{s})}{\rm tr}_{\xi}{\rm e}^{2m_{0}\beta_{s}\xi -
\beta_{s}\xi^{2}}\int Du_{1} \cdots \int Du_{Q-1}\{\xi_{i}-\Omega(\langle
\sigma_{i}^{\alpha}\rangle)\}^{2}.
\label{stat11}
\end{equation}

We next assume a condition to compare the performance of restoration between
using the bit-decomposed data and the $Q$-Ising form. The condition is that the
distance between the original and the degraded images in our method is
equal to that in the $Q$-Ising form. We can compare the difference of
performance in disparate noisy channels by this condition.
The distance between the original and degraded images in our method
$H_{{\rm D}}^{\tau}(\mbox{BDD})$ is expressed similarly to the above
expression. 
It turns out that the distance $H_{{\rm D}}^{\tau}(\mbox{BDD})$ becomes
simple in the thermodynamic limit $N \rightarrow \infty$. 
\begin{eqnarray}
H_{\rm D}^{\tau}(\mbox{BDD})
& = &
\frac{1}{N} \sum_{i} \{(\xi_{i,1}+\xi_{i,2}+ \cdots +\xi_{i,Q-1}) 
- (\tau_{i,1}+\tau_{i,2}+ \cdots + \tau_{i,Q-1})\}^{2}
\nonumber\\
& =  &
\bigg\langle \bigl(\sum_{k} \xi_{k}-\sum_{k}\tau_{k}\bigr)^{2}
\bigg\rangle_{\tau_{1},\cdots ,\tau_{Q-1},\xi_{1}\cdots \xi_{Q-1}}
\nonumber\\
&=&
(Q-1)\tau^{2} + \frac{(\tau_{0}-1)^{2}}{Z(\beta_{s})}
{\rm tr}_{\xi_{1} \cdots \xi_{Q-1}}{\Bigg [} \sum_{k=1}^{Q-1}\xi_{k}{\rm
e}^{2\beta_{s}m_{0}\sum_{k}\xi_{k}-\beta_{s}(\sum_{k}\xi_{k})^{2}} {\Bigg ]}
\nonumber \\
& = & (Q-1)\tau^{2}.
\label{stat14}
\end{eqnarray}
When the mean $\tau_{0}$ of the Gaussian noise is 1, the distance
$H_{{\rm D}}^{\tau}(\mbox{BDD})$ depends on the variance $\tau^{2}$
only. In the $Q$-Ising form,
$H_{{\rm D}}^{\tau}(\mbox{$Q$-Ising})$ is
\begin{eqnarray}
H_{\rm D}^{\tau}(\mbox{$Q$-Ising}) & = & \frac{1}{N}\sum_{i}
\{ \xi_{i} - \tau_{i} \}^{2}
\nonumber \\
& = &
{\tau^{\prime}}^{2} + \frac{(\tau'_{0}-1)^{2}}{Z(\beta_{s})}
{\rm tr}_{\xi}{\Bigg [} \xi{\rm e}^{2\beta_{s}m_{0}\xi-\beta_{s}\xi^{2}}
{\Bigg ]}\nonumber \\
&=& {\tau'}^{2}
\label{stat15}
\end{eqnarray}
where ${\tau'}^{2}$ is the variance of 
a Gaussian channel in the $Q$-Ising
form and is different from $\tau$ in 
the bit-decomposed data. 
When $\tau'_{0}=0$, $H_{\rm
D}^{\tau}(\mbox{$Q$-Ising})={\tau^{\prime}}^{2}$ 
similarly to the bit-decomposed data case.
As the rate of degradation is equal to each other, namely 
$H_{\rm D}^{\tau}(\mbox{BDD})=H_{\rm D}^{\tau}(Q\mbox{-Ising})$, 
the variance of a noisy channel between in our method (\ref{stat14}) and
that in the $Q$-Ising form (\ref{stat15}) satisfies the following relation.
\begin{eqnarray}
\tau = \frac{\tau'}{\sqrt{Q-1}}
\label{stat16}
\end{eqnarray}

Next, we calculate the restoration of gray-scale image when $Q=3$. We
plot the magnetization $m_{0}$ of original image as a function of source
temperature $T_{s}$ for $Q=3$ in FIG. \ref{fig4}. In high temperature
region $T_{s}\rightarrow \infty$, the magnetization becomes
$m_{0}=(0+1+2)/3=1$ since each spin takes all the values with the
same probability $1/3$. We see that two locally stable states
are generated in the middle range of temperature. The globally
stable state is the line of $m_{0}=1$. The transition temperature between the
paramagnetic phase and the ferromagnetic phase is $T_{c}\sim 0.9$.
These locally stable states become more stable with the decrease of
temperature and
correspond to the globally stable states $m_{0}=0$ and $m_{0}=2$ in
$T_{s}=0$. That is, the system of the ferromagnetic $Q$-Ising model for
$Q=3$ has triple degeneracy.

We use a snapshot of the system when the magnetization is
$m_{0}=1$ at $T_{s}=0.75$ as the original image. When the distance
between the original image and the degraded image is 
$H_{{\rm D}}^{\tau}=1.0$, macroscopic quantities behave as 
in FIG. \ref{fig5}.
The magnetization $m$ of equation (\ref{stat10}) at
the ratio of hyper-parameters $H\equiv h/\beta_{m}=0.25$ has
three stable states at $T_{m}=0$ as $m_{0}$ we see in the original
image. At low temperature, the
system has one globally stable state $m=1$ and two locally stable states
$m=0$ and $m=2$. However these locally stable states vanish as the ratio
of hyper-parameters $H$ becomes large. Seeing the free energy in FIG. 
\ref{fig5}, we find that the state of system branches out into a global
minimum and local minima at $T_{m}=0.27$.

The mean square error $H_{{\rm D}}$ using
$m$ of the global minimum is smaller than that using $m$ of local minimum
in all temperature region and gives the optimal value for the ratio of 
hyper-parameters $H$.
The mean square error $H_{{\rm D}}$ between the $m$ of global minimum and
the original image $m_{0}$ in some ratio of hyper-parameters $H$ are
described in FIG. \ref{fig6}. 
The smallest mean square error is given at $T_{m}=0.75$ and
$H=0.75$. When the distance between the original and degraded
images is 1.0 $(H_{{\rm D}}^{\tau}=1.0)$, we see that the deviation $\tau$ of
Gaussian noise becomes $1/\sqrt{Q-1}=1/\sqrt{2}$ by equation
(\ref{stat14}). Therefore, when the ratio of hyper-parameters
corresponds to that of the prior parameters
$(H=h/\beta_{m}=(\tau_{0}/2\tau^{2})/\beta_{s} =0.75)$, the
optimal restoration is given at the temperature corresponding to the source
temperature $(T_{m}=T_{s}=0.75)$ similarly to the Ising model.
At high temperatures $T_{m}\rightarrow \infty$, all states appear with equal
probability. Then the thermal average of a spin $\sigma$ becomes
$\langle \sigma \rangle = (0+1+2)/3=1$ at all pixels. The mean
square error is
\begin{equation}
H_{{\rm D}}(T_{m}\rightarrow
\infty)=\frac{\sum_{\xi=0}^{3}(\xi-1)^{2}
{\rm e}^{2m_{0}\beta_{s}\xi-\beta_{s}\xi^
{2}}}{\sum_{\xi=0}^{3}{\rm e}^{2m_{0}\beta_{s}\xi-\beta_{s}\xi^{2}}}\sim 0.3452.\label{stat17}
\end{equation}
If the distance between the original and degraded images, 
$H_{{\rm D}}^{\tau}(\mbox{BDD})$ or 
$H_{{\rm D}}^{\tau}(\mbox{$Q$-Ising})$, is extremely large, the
performance of restoration does not become smaller than this value.

We compare the optimal performance of restoration using our method with
that using the $Q$-Ising form under the condition that the degradation
of image in our method is equivalent to the one in the $Q$-Ising form,
namely $H_{{\rm D}}^{\tau}\,\,(\mbox{BDD})=
H_{{\rm D}}^{\tau}\,\,(\mbox{$Q$-Ising form})$. 
In the $Q$-Ising form case, the deviation is $\tau =1.0$ by equation
(\ref{stat15}) when $H_{{\rm D}}^{\tau}=1.0$.
The optimal restoration in the $Q$-Ising form is also given in the
ratio of hyper-parameters corresponding to that of source parameters
$(H= h/\beta_{m} = (\tau_{0}/2\tau^{2})/\beta_{m} =0.375)$.
The optimal performance of restoration in both our method and the
$Q$-Ising form is shown in FIG.  \ref{fig6}. We see that the
restoration by our method gives better performance than that of the
$Q$-Ising form.

\section{Monte Carlo simulation}
In this section, we investigate realistic pictures in two dimensions
and check the results obtained in the infinite-range model. It is, 
however, difficult to investigate the restoration of real images by
analytical methods. We discuss the restoration of two-dimensional images
by means of Monte Carlo simulations.

We suppose that the original image is generated by the Boltzmann
probability distribution as we did in the infinite-range model case and we
use a snap shot of the $Q$-Ising model at a temperature as the original
image. 

In digital images, the pixel takes discrete values. 
Even if real-valued Gaussian noise is added in the degradation process, 
we have to regard such a noise as the discrete noise (for example,
binary noise) when both the original and the degraded images are
digital. (This is essentially the same situation as in a
error-correcting codes. \cite{bib24}) Therefore, we consider the
binary noise caused by the binary symmetric channel with the error
probabilities $p=0.10\,\,(Q=4)$ and $p=0.15\,\,(Q=8)$, which correspond
to the parameters $\beta_{\tau}\sim 2.2$ and $\beta_{\tau}\sim 1.7$ by
Eq. (\ref{ir6}).  
The size of digital image is $100\times100$ and averages over ten
samples are taken at each data point.

FIG. \ref{fig-mc3} (the upper figure) 
represents the mean square error of several $H$
when $Q=4,\,\,T_{s}=0.35,\,\,m_{0}\sim 1.2,\,\,p=0.10$ and 
$H_{{\rm D}}^{\tau}\sim 0.30$.
The optimal restoration is given at
$T_{m}=T_{s},\,\,H=h/\beta_{m}=\beta_{\tau}/\beta_{s}
\sim 0.75$ like in the infinite-range model. 

The mean square errors for $Q=8,\,\,T_{s}=0.4,(0.6),\,\,m_{0}\sim
2.2,(3.0),\,\,p=0.15,\,\,H_{{\rm D}}^{\tau}\sim 1.0$ are 
also shown in FIG. \ref{fig-mc3} (the lower two figures.). 
We see that the best performance of restoration at each
temperature is obtained at the ratio of hyper-parameters $H$
corresponding to that of the prior parameters, and the performance of
restoration in $T_{s}=0.4$ (the lower left) is better 
than that in $T_{s}=0.6$ (the lower right) as written
before.

\section{Mean-field annealing}
The amount of computation required for restoration by
means of Monte Carlo simulation is enormous. Hence, in the present 
section, we reconstruct the original image from binary 
degraded images by means of the mean-field annealing 
\cite{bib25,bib26} with periodic boundary conditions. 
This method enables us to search for the optimal solution quickly. 
We apply the mean-field approximation to the posterior probability
distribution. 
\begin{eqnarray}
\lefteqn{P(\{\mbox{\boldmath $\sigma$}\}|\{\mbox{\boldmath $\tau_{1}$}\},
\{\mbox{\boldmath $\tau_{2}$}\},\cdots \{\mbox{\boldmath
$\tau_{Q-1}$}\})}\nonumber \\
&=&
\frac{1}{Z(\beta_{m},h)}{\exp}\left(
-h\sum_{i}\sum_{k}(\sigma_{i,k}-\tau_{i,k})^{2}-
\frac{\beta_{m}}{2N}\sum_{i,j}(\sigma_{i}-\sigma_{j})^{2}
\right)
\label{mf1}
\end{eqnarray} 
In order to treat each site separately, we trace out with respect to all
pixels besides $i$-th pixel in the above probability distribution.
\begin{eqnarray}
\rho_{ij}(n) & = & \prod_{kl \neq ij}{\rm tr}_{\sigma_{1},\cdots,\sigma_{Q-1}}
P(\{\mbox{\boldmath $\sigma$}\}|\{\mbox{\boldmath $\tau_{1}$}\},
\{\mbox{\boldmath $\tau_{2}$}\},\cdots \{\mbox{\boldmath $\tau_{Q-1}$}\})
\nonumber \\
& = & {\rm Tr}_{\sigma_{1},\cdots,\sigma_{Q-1}}
P(\{\mbox{\boldmath $\sigma$}\}|\{\mbox{\boldmath $\tau_{1}$}\},
\{\mbox{\boldmath $\tau_{2}$}\},\cdots \{\mbox{\boldmath $\tau_{Q-1}$}\})
\delta(n,\sigma_{ij})
\label{mf2}
\end{eqnarray}
where $(ij)$ represents the site index in two dimensions: $i$
and $j$ are the coordinates of $x$ and $y$ axes, respectively.
This expression is called the marginal probability
distribution. According to mean-field approximation, the posterior
probability distribution (\ref{mf1}) is approximated by the
product of the marginal probability distribution (\ref{mf2})
as follows. 
\begin{equation}
P(\{\mbox{\boldmath $\sigma$}\}|\{\mbox{\boldmath $\tau_{1}$}\},
\{\mbox{\boldmath $\tau_{2}$}\},\cdots \{\mbox{\boldmath $\tau_{Q-1}$}\})
\simeq
\prod_{(ij)}\rho_{ij}(n).
\label{mf3}
\end{equation}

The recursion relation for the iterative algorithm is derived from a
variational principle. To derive the recursion relation, we substitute
the expression of approximation (\ref{mf3}) to the free energy.
\begin{equation}
F(\rho) = E(\rho) - T_{m}S(\rho).
\label{mf4}
\end{equation}
The variation of the above free energy is calculated with respect to the
marginal probability distribution at a certain site: $\rho_{ij}$.
Finally, the recursion relation with respect to the local magnetization
for the iteration algorithm is obtained as 
\begin{eqnarray}
m_{ij}^{t+1}=\frac{\sum_{\sigma_{ij}=0}^{Q-1}\sigma_{ij}\exp(H_{{\rm
MF}}^{t})}{Z_{{\rm MF}}},
\label{mf5}
\end{eqnarray}
\begin{eqnarray}
H_{{\rm MF}}^{t} & = &
\frac{\beta_{m}}{2}(m_{i,j+1}^{t}+m_{i,j-1}^{t}+m_{i+1,j}^{t}+m_{i-1,j}^{t})
\sigma_{ij}\nonumber\\
& & -\beta_{m}\sigma_{ij}^{2}-h\sum_{k}(\sigma_{ij,(k)})^{2}
+2h\sum_{k}\tau_{ij,(k)}\sigma_{ij,(k)},
\label{mf6}
\end{eqnarray}
where $Z_{{\rm MF}}$ is the normalization constant in the mean-field
approximation. \\
We solve the above relations numerically under the convergence condition 
\begin{eqnarray}
\epsilon^{(t)}\equiv \frac{1}{N}
\sum_{(ij)}^{N}|m_{ij}^{(t+1)}-m_{ij}^{(t)}|<10^{-8},
\label{mf7}
\end{eqnarray}
and obtain approximately the restored image at each temperature. 
The annealing schedule is set at $\Delta T_{m}=0.01$.

We compare the performance of restoration using our method with that using the
$Q$-Ising form for five kinds of standard images \cite{bib}. We choose $Q=8$,
system size=$200\times200$ and $H_{{\rm D}}^{\tau}\simeq 1.00$. 
The result is shown in table \ref{tab-mf1}. 
The original versions of these five standard 
image are shown in FIG. \ref{five}.  

The original, degraded and
optimal restored images in our method and
$Q$-Ising form are shown in FIG.  \ref{stand4}  
for the case of "lena".

In TABLE \ref{tab-mf1}, the ratio of pixels at which the nearest
neighboring pixels take the same value is also represented. We see that
the performance of restoration using our method is better than that
using the $Q$-Ising form in the standard pictures ``chair'' and
``girl''. However, the $Q$-Ising form is better in ``house'', ``lena'' and
``mandrill''. One of the differences is that ``chair'' and ``girl'' have a
lot of large parts and do not have short edges compared with the
other pictures as we see intuitively. 
In TABLE \ref{tab-mf1}, that ``chair'' has obviously large parts is 
quantitatively expressed by the ratio of pixels at which the nearest
neighboring pixels take the same value. 
However, the ratio in ``girl'' is as much as that in ``house'' and
``lena''.

In order to investigate this aspect in more detail, we show the ratio of
pixels at which the difference among the value of the nearest neighboring
pixels is smaller than $2$ in table \ref{tab-mf2}. 
As seen in TABLE \ref{tab-mf2},
``girl'' is similar to ``chair'' rather than ``house'' and ``lena''. 
The effects of noise in the bit-decomposed data and $Q$-Ising form are
clarified by comparing the NNP1$-$D in TABLE  \ref{tab-mf1} and
NNP2$-$D in TABLE  \ref{tab-mf2}. NNP1$-$D in our method is smaller
than that in the $Q$-Ising form for all kinds of standard images, but
the decrease of NNP2 from the original image to the degraded image in
our method is small compared with that in the $Q$-Ising form for five
standard images. That is, the noise in our method affects the original
image widely, but the shift of value in a site is small. On the other
hand the shift of value in a site is large in the $Q$-Ising form.
Therefore the noise in our method has the effect of smoothing, and the
restoration using our method is efficient for original images in
which eliminating noise is more significant rather than preserving
informations of original image as ``chair'' and ``girl''.

Furthermore, we compare the performance using our method and
in the $Q$-Ising form for the $H_{{\rm D}}^{\tau}\sim 2.0$ case. Other
conditions are the same as the previous case. We can simply investigate the
tolerance of our method and $Q$-Ising form against noise by comparing
$H_{{\rm D}}^{\tau}\sim1.0$ and $2.0$ cases. The performance
of restoration is shown in TABLE  \ref{tab-mf3} and 
the resultant image is presented in FIG.  \ref{stand9} 
for the case of "lena".
The performance of restoration is clearly worse than that in the
previous case for all pictures, and the difference of performance between
 our method and the $Q$-Ising form is clear. Moreover, the restoration
 using our method gives better performance than that using the
 $Q$-Ising form in standard image ``house''. The result is  different
 from that when $H_{{\rm D}}^{\tau}\sim 1.0$. In the $Q$-Ising form, 
large clusters remain in the restored image as in FIG. 
\ref{stand9}. Accordingly we conclude that the restoration using the
bit-decomposed data is not affected by the noise compared with that
using the $Q$-Ising form in this case.

Consequently we found that the restoration using our method is more
efficient than that in the $Q$-Ising form in original images where the
nearest neighboring pixels take the close values. The reason is that to
eliminate the added noise is more significant than the information of
the original image in such images. Regardless of the $Q$-Ising form and
our method, the restoration of such images tend to give good performance
compared with that of other images in the image restoration using the
method of statistical mechanics.
Moreover, our method can give the adequate restored image in original
images ``chair'', ``girl'' and ``house'' even if the noise is strong to
a certain extent: for example, $H_{{\rm D}}^{\tau}\sim 2.0$.

Finally, we try to restore the original image by using the composition
of our method and the $Q$-Ising form.
Two processes are considered in restoration using the composition. 
One is the method of restoration where the original image is
restored from $Q-1$ sets of binary degraded images which are generated
from the received, degraded $Q$-value image by the method of threshold
division. Another is the method of restoration where the
original image is restored by using the $Q$-Ising form after translating
$Q-1$ sets of binary degraded images into a $Q$-value degraded image.
These restoration processes are shown in FIG.  \ref{fig-qb}.

The second process is not efficient in the restoration of gray-scale image
because the performance of restoration is worse than that both using our
method and the $Q$-Ising form. On the other hand, the first process gives
interesting results as in TABLE \ref{tab-qb} when $H_{{\rm D}}^{\tau}\sim1.0$.
Seeing in TABLE \ref{tab-qb}, we find that this method gives
better performance than our method and the $Q$-Ising form for all
standard images: ``chair'', ``girl'', ``house'', ``lena'' and ``mandrill''.
The restoration of standard images using this method 
for the case of "lena" is shown 
in FIG.  \ref{lena-qb}. 
This idea that, after a $Q$-value data are received, they are
decomposed into binary data has been already used in the field of
engineering. The filter is called the {\it Stack filter} 
\cite{pitas90} which works on
decomposed binary data after reception.

The Stack filter by means of the method of statistical mechanics as the
first process gives better performance than the other two methods, but the
theoretical framework is not clear. Because we have to assume the noisy
process matching the degraded data in the Bayesian viewpoint, it is very
difficult for us to estimate the noisy channel matching such binary data.   
Accordingly, it is well possible that the hyper-parameter $h$
can not be estimated appropriately for the hyper-parameters estimation
in the Stack filter by means of the method of statistical mechanics.

\section{Summary and discussions}
In this paper, we investigated the restoration of gray-scale image
using the bit-decomposed data instead of the conventional $Q$-Ising form
and found the conditions that our method is more efficient than the
method of the $Q$-Ising form.

In the infinite-range model, we analyzed the image
restoration when the original image is affected by the Gaussian noise,
and obtained the static properties of image restoration.
In order to obtain the averaged performance, we calculated the restoration of
gray-scale image by means of the replica method. Then we found that the mean
square error has a minimum at the finite temperature for any ratio of the
hyper-parameters $H$. However, when the noise rate is extremely large, 
the minimum sinks under the mean square error 
$H_{{\rm D}}(T_{m}\rightarrow \infty)$ in the high temperature
limit. In other words, we can only obtain the restored image in
which all states appear with equal probability as the optimal restored
image when the original image is very corrupted.
The best performance of restoration is given at the temperature
at which the original image is generated, when the ratio of
hyper-parameters corresponds to that of the prior parameters. 
This result is also common between the Ising spin and the $Q$-Ising spin
cases. Furthermore, we obtained the quantitative result that the
performance of restoration using our method is better than that using
the $Q$-Ising form when 
$H_{{\rm D}}^{\tau}(\mbox{BDD})=H_{{\rm D}}^{\tau}(\mbox{Q-Ising})$.

We analyzed the restoration of realistic images in two dimensions by
means of Monte Carlo method and the mean-field annealing approximately.
Using Monte Carlo method, we confirmed the result of the infinite-range
model that the mean square error gives the optimal minimum value at the
temperature corresponding to the source temperature when the ratio of
hyper-parameters corresponds to prior parameters for a snap-shot of
the $Q$-Ising model for $Q=4$ and $Q=8$.
We restored five standard images by means of the mean-field annealing
and found that the restoration using our method is better than
that using the $Q$-Ising form in the original images where
keeping the information of original image is not significant compared
with the eliminating the added noise as ``chair'' and ``girl''. 
The reason is that the effects of noise in our method contain the
effect of smoothing. However, our method is not efficient in images
which need information of the original image strongly.

Thus there are simply two types in natural images. One is the
image which consists of a little long edges and large surfaces, and
those images depend on a basic and common property of natural image
strongly. Another is the image which has many short edges and small
clusters, and it is difficult to distinguish whether the original or
degraded images in such images. Therefore not only corrupted parts
but correct parts may be destroyed by the effect of smoothing in such images.
In restoration of images by means of the method of statistical mechanics,
the effect of smoothing is strong compared with keeping the information of
original image that the degraded image contains because we use it as
common property of natural images. Therefore, we need some additional
information on the original image.

Furthermore, we found also that the restoration using our method is not
affected by noise compared with that using the $Q$-Ising form.      
One of the reasons is due to that the output, namely the degraded data, 
takes the close value to input data by the effects of noise in our
method. This is the most significant property of our method.
We expect that the restoration using our method gives further better
performance when $Q$-value is large.
Because we could not distinguish visually the difference between close
values with increasing the gray-scale level ($Q$-value), the effect of
noise may be suppressed visually.
Accordingly, we may be obtain the restored image that looks closer to
the original image intuitively.
\\

The authors thank Professor Hidetoshi Nishimori for useful
discussions.

\begin{table}[h]
\caption{Comparison of performance between our method
 and the $Q$-Ising form. The ratios of pixels at which the nearest
 neighboring pixels take the same value in the original image, the
 degraded image and restored image are represented as 'NNP1'.}
\label{tab-mf1}
\end{table}

\begin{table}[h]
\caption{Ratio of pixel at which the difference among the value of 
the nearest neighboring pixels is smaller than 2 in the original,
 degraded and restored images. }
\label{tab-mf2}
\end{table}

\begin{table}[h]
\caption{Comparison of performance between using our method and
 the $Q$-Ising form for five standard images when 
$H_{{\rm D}}^{\tau}\sim 2.0$. 
In this case, the restoration using our method is more
efficient than that using the $Q$-Ising form in three standard images:
"chair", "girl" and "house". }
\label{tab-mf3}
\end{table}
\begin{table}
\caption{Comparison of performance among using our method, the $Q$-Ising 
form and the composition for five standard images 
when $H_{{\rm D}}^{\tau} \sim 1.0$. In this case, the restoration 
using the composition is more efficient than that using our 
method and the $Q$-Ising form in all standard images. 
}
\label{tab-qb}
\end{table}


\begin{figure}
\caption{The example of decomposition by means of the 
 method of threshold division in the $Q=3$ and $6$-site case.}
\label{fig2}
\end{figure}

\begin{figure}
\caption{Magnetization of the original
 image as a function of the source temperature for $Q=3$. The solid line
 corresponds to a globally stable solution $m_{0}=1$ and the dotted lines
 correspond to locally stable solutions $m_{0}=0$ and $m_{0}=1$.}
\label{fig4}
\end{figure}

\begin{figure}
\caption{The magnetization $m$ and corresponding free energy $f_{\rm RS}$ are plotted in the upper-left and upper-right figures. The mean square error is plotted in the lower-left figure. The lower-right figure represents an enhancement of the mean square error around the optimal values. The solid line corresponds to the globally stable solution.}
\label{fig5}
\end{figure}

\begin{figure}
\caption{Left figure is the mean square error as a function 
of temperature for several $H$. 
The solid line corresponds to the 
optimal value,  $T_{m}=T_{s}$. Right figure 
represents comparison between the mean
square error in the bit-decomposed data and the $Q$-Ising form for the
optimal hyper-parameters $H=H^{\rm opt}$. The solid line is that for
the BDD case and the dotted line is the $Q$-Ising case.}
\label{fig6}
\end{figure}

\begin{figure}
\caption{The mean square error in the $Q=4$ case is calculated by
 Monte Carlo simulation (the upper figure). 
 Averages over ten samples that (system size is $100\times100$) 
 is taken at each data point. The ratio of
 hyper-parameters $H$ is chosen to be $H=0.6$, $0.75$ (optimal) and $1.0$. 
 The mean square error when  $Q=8,\,p=0.15,\,H_{{\rm D}}^{\tau}\sim1.0$ 
 (the lower figures). 
 The lower left figure represents 
 the mean square error in $T_{s}=0.4$, and the ratio
 of hyper-parameters $H$ is chosen to be $H=0.6$, $0.68$ (optimal) and $0.8$. 
 The lower right figure represents 
 the mean square error in $T_{s}=0.6$, and the ratio of 
 hyper-parameters $H$ is chosen to be $H=0.8$, $1.0$ (optimal) and $1.2$.}
\label{fig-mc3}
\end{figure}

\begin{figure}
\caption{Standard images: 
"chair", "girl", "house", "lena" and "mandrill". }
\label{five}
\end{figure}

\begin{figure}
\caption{Standard image ``lena''. Left side figures represent the
restoration using the bit-decomposed data, and right side figures represent
the restoration using the $Q$-Ising 
form for $H_{{\rm D}}^{\tau}\sim 1.0$. } 
\label{stand4}
\end{figure}

\begin{figure}
\caption{Standard image ``lena''. Left side figures represent the
 restoration using the bit-decomposed data, and right side figures represent
 the restoration using the $Q$-Ising form for $H_{{\rm D}}^{\tau}\sim 2.0$. }
 \label{stand9}
\end{figure}

\begin{figure}
\caption{Restoration processes using the composition of both our
 method and the $Q$-Ising form. In the upper process (first process), we
 decompose a $Q$-value degraded image into $Q-1$ sets of binary degraded
 images after  we received a $Q$-value degraded image. 
 Using the binary degraded images, we restore the original image. 
 On the other hand, in the lower process (second process), we received $Q-1$
 sets of binary degraded images. After they are translated into a
 $Q$-value degraded image, we restore the original image using the
 $Q$-Ising form. }
\label{fig-qb}
\end{figure}

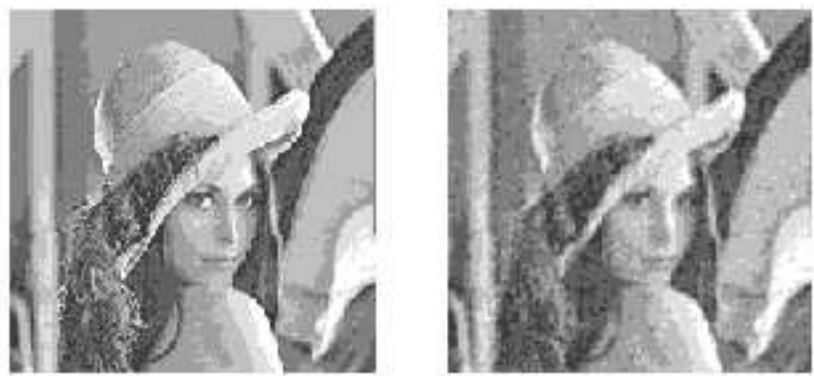
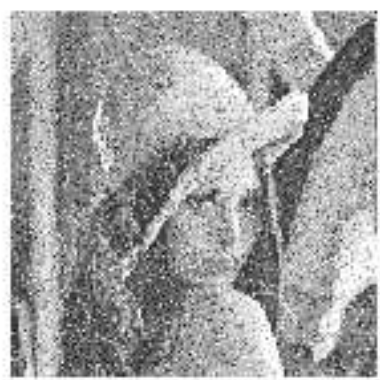
\begin{figure}
\caption{Standard image ``lena''. The left, center and right figures
 are the original, degraded and restored images, respectively, in the
 composition process when $H_{{\rm D}}^{\tau}\sim1.0$.}
\label{lena-qb}
\end{figure}


\begin{thebibliography}{l}
\bibitem{bib}The standard images which are used in this paper are
available at {\sf ftp://ftp.lab1.kuis.kyoto-u.ac.jp/pub/sidba/}.
\end{thebibliography}
\end{document}